\documentclass[pteplogo]{ptephy_v2}
\preprintnumber{} 
\usepackage{hyperref}

\newcommand{\im}{\mathrm{i}}  
\newcommand{\e}{\mathrm{e}}  
\newcommand{\dif}{\mathrm{d}}  
\newcommand{\del}{\partial}  

\newcommand{\bs}{\boldsymbol}
\newcommand{\fr}{\frac}

\newcommand{\lb}{\left}
\newcommand{\rb}{\right}
\newcommand{\mcl}{\mathcal}

\newcommand{\bmx}{\left(\begin{matrix}}
\newcommand{\emx}{\end{matrix}\right)}

\newcommand{\nom}{\nonumber \\}

\begin{document}

\title{Anomalous Dynamical Screening of Relativistic Plasma in a Magnetic Field}

\author{Sota Hanai}
\affil{Department of Physics, Keio University,\\
3-14-1 Hiyoshi, Yokohama, Japan \email{souta.h.0619@keio.jp}}

\begin{abstract}
We study collective excitations in a relativistic collisionless plasma composed of massless fermions subject to an external magnetic field.
We include dynamical electromagnetism and fluctuations of the chiral charge, while maintaining a vanishing net chiral charge.
Within the framework of chiral kinetic theory, we find that the chiral anomaly gives a correction to the transverse photon self-energy and that a novel type of dynamical screening can emerge.
In the strong magnetic field limit, we also show that the collective mode has a gap using the lowest Landau level approximation.
We further discuss implications of the anomalous dynamical screening for neutron star phenomenology.
\end{abstract}

\subjectindex{}

\maketitle

\section{Introduction}
\label{sec:Introduction}
Chirality plays a crucial role in relativistic plasmas, such as those found in heavy-ion collisions \cite{Huang:2015oca,Kharzeev:2015znc}, neutron stars \cite{Kamada:2022nyt}, and the early Universe \cite{Kamada:2022nyt}.
In many of these systems, the plasma can coexist with strong magnetic fields.
A prototypical phenomenon induced by a magnetic field is the chiral magnetic effect (CME) \cite{Vilenkin:1980fu,Nielsen:1983rb,Alekseev:1998ds,Fukushima:2008xe}, where an electric current is induced along the magnetic field in the presence of chirality imbalance.
The CME gives rise to various collective excitations.
In a background magnetic field, a density wave called the chiral magnetic wave (CMW) propagates along the magnetic field.
The CMW can occur even in the absence of a net chirality imbalance due to the chiral separation effect (CSE) \cite{Son:2004tq,Metlitski:2005pr}, where fluctuations of the number density can induce fluctuations of the chiral charge density.
In plasmas with chirality imbalance, the collective mode of dynamical electromagnetic fields exhibits an instability called the chiral plasma instability (CPI) \cite{Akamatsu:2013pjd,Akamatsu:2014yza}.
The collective modes of magnetized relativistic plasmas, including dynamical electromagnetic fields, have been investigated within the framework of hydrodynamics \cite{Rybalka:2018uzh,Shovkovy:2018tks}.

Collective excitations in relativistic plasmas have been studied not only in the hydrodynamic regime but also in the collisionless regime \cite{Gorsky:2012gi,Stephanov:2014dma}.
The previous studies investigated the zero sound of chiral fermions at finite chiral chemical potential in a non-dynamical background magnetic field, without taking the chiral charge fluctuation into account.
On the other hand, one can also consider the fluctuations of the chiral charge and electromagnetic fields in a chirally symmetric plasma.
In particular, due to the dynamical electric field, the chiral charge in a background magnetic field is affected by the chiral anomaly \cite{Adler:1969gk,Bell:1969ts}.

In this paper, we study the collective excitations of a relativistic magnetized plasma using chiral kinetic theory \cite{Son:2012wh,Stephanov:2012ki,Son:2012zy}.
We derive the dispersion relations for general frequencies and wavenumbers, including the dynamical electromagnetic fields and the chiral charge fluctuation.
We then show that, due to the back reaction induced by the chiral anomaly, the transverse photon self-energy is modified, and that dynamical screening distinct from the conventional Landau damping can appear.
We also examine the collective mode in the lowest Landau level (LLL) approximation.
In addition, we discuss implications for transport properties in neutron stars.

This paper is organized as follows.
In Sect.~\ref{sec:EM_mag}, we review electrodynamics in media with a background magnetic field.
In Sect.~\ref{sec:Collective_modes}, we derive the dispersion relations of the collective excitations.
In Sect.~\ref{sec:Imp_neutron_star}, we discuss the implications of our results for neutron star phenomenology.
Our conclusions and outlook are presented in Sect.~\ref{sec:Conclusion_Outlook}.

Throughout this paper, we adopt natural units with $\hbar = c = k_{\rm B} = \varepsilon_0 = \mu_0 = 1$.

\section{Electrodynamics in a background magnetic field}
\label{sec:EM_mag}
In this section, we first review the electrodynamics in a magnetized medium (see, e.g., Ref.~\cite{Landau:8}).
The argument here is independent of the microscopic details of the medium.
To distinguish the dynamical electromagnetic fields from the external ones, we write
\begin{align}
    \bs{E}=\delta\bs{E}\,,
    \qquad
    \bs{B}=\bs{B}_{\rm ex}+\delta\bs{B}\,,
\end{align}
where $\bs{B}_{\rm ex}$ is a uniform external magnetic field.
Without loss of generality, we set a uniform external magnetic field along the $z$-axis ($\bs{B}_{\rm ex}=B_{\rm ex}\bs{e}_z$).

The dynamics of electromagnetic fields in media are governed by the Maxwell equations:
\begin{align}
    \bs{\nabla}\times\bs{E}=-\del_t\bs{B}\,,
    \qquad
    \bs{\nabla}\times\bs{B}=\del_t\bs{E}+e\bs{j}\,,
\end{align}
where $e\bs{j}$ is the induced electric current.
In the following, we assume that there are no external electric sources and write $\bs{j}=\delta\bs{j}$.
To investigate the collective behavior, we assume a solution in the form of $\delta\bs{E},\delta\bs{B}\propto\e^{-\im(\omega t-\bs{k}\cdot\bs{x})}$, and the Maxwell equations are rewritten as
\begin{align}
\label{eq:Maxwell_eqs_Fourier}
    \bs{k}\times\delta\bs{E}=\omega\delta\bs{B}\,,
    \qquad
    \bs{k}\times\delta\bs{B}=-\omega\delta\bs{E}-\im e\delta\bs{j}\,.
\end{align}
In the linear response theory, the electric current can be expressed as $e\delta\bs{j}=-\im\omega\delta\bs{P}$ where $\bs{P}$ is the polarization vector.
Thus, we define the permittivity tensor $\varepsilon$ as
\begin{align}
\label{eq:Permittivity_current_general}
    (\varepsilon-1)\delta\bs{E}
    =
    \im\fr{1}{\omega}e\delta\bs{j}\,.
\end{align}
The details of the medium are included in the permittivity.
We will derive the permittivity of the relativistic plasma in the next section.
Eliminating $\delta\bs{B}$ from Eq.~(\ref{eq:Maxwell_eqs_Fourier}), we obtain 
\begin{align}
\label{eq:Maxwell_media_r}
    \bs{r}^2\delta\bs{E}-(\bs{r}\cdot\delta\bs{E})\bs{r}-\varepsilon\delta\bs{E}
    =
    0\,,
\end{align}
where we defined a refractive vector as $\bs{r}\equiv\bs{k}/\omega$.
Since the refractive index is generally a complex value, we write $\bs{r}^2$ instead of $|\bs{r}|^2$.
The dispersion relation can be derived by solving this equation. 

Although the dynamics of the electromagnetic fields generally depends on the permittivity, one can find general properties of the dispersion relation by symmetry.
In the presence of the background magnetic field, the permittivity tensor is invariant under the $\fr{\pi}{2}$-rotation about the $z$-axis, and we have $R\varepsilon R^{-1}=\varepsilon$ where $R$ is a rotation matrix,
\begin{align}
    R=
    \lb(
    \begin{array}{ccc}
        0 & -1 & 0  \\
        1 & 0 & 0 \\
        0 & 0 & 1
    \end{array}\rb)\,.
\end{align}
Hence, the permittivity tensor is characterized by three parameters as
\begin{align}
    \varepsilon
    =
    \lb(
    \begin{array}{ccc}
        \varepsilon_{xx} & \varepsilon_{xy} & 0  \\
        -\varepsilon_{xy} & \varepsilon_{xx} & 0 \\
        0 & 0 & \varepsilon_{zz}
    \end{array}\rb)\,.
\end{align}
The off-diagonal components are antisymmetric.
According to Onsager's theorem, the permittivity tensor satisfies $\varepsilon^{ij}(B_{\rm ex})=\varepsilon^{ji}(-B_{\rm ex})$, implying that the off-diagonal components are odd in the magnetic field, while the diagonal components are even.

As will be shown later, the electromagnetic waves propagating perpendicular to the magnetic field are subject to the chiral anomaly.
In the following discussion, we take the propagation direction along the $x$-axis $(\delta\bs{E}\propto\e^{-\im(\omega t-k_xx)})$.
In this case, Eq.~(\ref{eq:Maxwell_media_r}) becomes
\begin{align}
    \lb(
    \begin{array}{ccc}
        -\varepsilon_{xx} & -\varepsilon_{xy} & 0 \\
        \varepsilon_{xy} & r_x^2-\varepsilon_{xx} & 0 \\
        0 & 0 & r_x^2-\varepsilon_{zz}
    \end{array}\rb)
    \lb(
    \begin{array}{c}
        \delta E_x \\
        \delta E_y \\
        \delta E_z 
    \end{array}\rb)
    =0\,.
\end{align}
From this equation, we obtain the dispersion relations of the two modes as
\begin{align}
\label{eq:Dispersion_relation_general}
    r_{x}^2=\varepsilon_{zz}\,,
    \qquad
    r_{x}^2
    =
    \varepsilon_{xx}+\fr{\varepsilon_{xy}^2}{\varepsilon_{xx}}\,.
\end{align}
The first electromagnetic wave is referred to as the ordinary mode (O-mode), and the second one is called the extraordinary mode (X-mode).
In quantum theory, the parallel component of the electric field gives rise to the chiral anomaly.
Therefore, the O-mode is modified by the quantum effect.

\section{Collective modes of relativistic collisionless plasma}
\label{sec:Collective_modes}
In this section, we discuss the collective behavior of the O-mode taking the chiral anomaly into account.
We consider the weak and strong magnetic field limits, respectively.
In the weak field case, the chiral kinetic theory is applicable, whereas we employ the LLL approximation in the latter limit.
Throughout this paper, we assume that the typical energy scales of the plasma, such as temperature and density, are sufficiently larger than the fermion mass, and the fermions are nearly gapless.
In general, the fermion mass gives rise to corrections, but we can neglect them in the collisionless regime, as discussed later.

\subsection{Chiral kinetic theory}
For completeness, we first review the chiral kinetic theory \cite{Son:2012wh,Stephanov:2012ki,Son:2012zy} (see also Refs.~\cite{Hidaka:2022dmn,Kamada:2022nyt} for recent reviews).
Kinetic theory is generally appropriate for studying non-equilibrium phenomena, not only when collisions are relevant but also when collisions are irrelevant.
The chiral fermions are treated semi-classically and described by the distribution function $f_{\lambda}(t,\bs{x},\bs{p})$ where $\lambda=\pm$ denotes the chirality: $\lambda=+$ for the right-handed and $\lambda=-$ for left-handed fermions. 
We also define the distribution function of the antifermion as $\bar{f}_{\lambda}(t,\bs{x},\bs{p})$.
In the following, we focus on fermions.
The distribution function for antifermions can be obtained by replacing $\lambda\to-\lambda$, $e\to-e$, and $\mu_{\lambda}\to-\mu_{\lambda}$, where $\mu_{\lambda}$ is the chemical potential for each chirality.

The Boltzmann equation in the collisionless regime is
\begin{align}
\label{eq:Boltzmann_collisionless}
    \fr{\del f_{\lambda}}{\del t}
    +\dot{\bs{x}}\cdot\fr{\del f_{\lambda}}{\del\bs{x}}
    +\dot{\bs{p}}\cdot\fr{\del f_{\lambda}}{\del\bs{p}}
    =0\,.
\end{align}
The crucial property of the chiral kinetic theory is encoded in the Berry curvature \cite{Berry:1984jv}, which corresponds to the ``magnetic field'' in momentum space.
For the chiral fermions, the Berry curvature is given by
\begin{align}
    \bs{\Omega}_{\lambda}
    =
    \lambda\fr{\bs{p}}{2|\bs{p}|^3}\,.
\end{align}
This has the same expression as the magnetic field induced by a magnetic monopole.
The equation of motion for the chiral fermions is modified as
\begin{align}
\label{eq:EoM_x}
    &\sqrt{G}\dot{\bs{x}}
    =
    \tilde{\bs{v}}+e\tilde{\bs{E}}\times\bs{\Omega}_{\lambda}+(\tilde{\bs{v}}\cdot\bs{\Omega}_{\lambda})e\bs{B}\,,
    \\
\label{eq:EoM_p}
    &\sqrt{G}\dot{\bs{p}}
    =
    e(\tilde{\bs{E}}+\tilde{\bs{v}}\times\bs{B})+e^2(\tilde{\bs{E}}\cdot\bs{B})\bs{\Omega}_{\lambda}\,,
\end{align}
where
\begin{align}
    \sqrt{G}
    \equiv
    1+e\bs{B}\cdot\bs{\Omega}_{\lambda}\,,
    \qquad
    \tilde{\bs{v}}\equiv\fr{\del\varepsilon_{\bs{p}}}{\del\bs{p}}\,,
    \qquad
    e\tilde{\bs{E}}\equiv e\bs{E}-\fr{\del\varepsilon_{\bs{p}}}{\del\bs{x}}\,,
\end{align}
with $\varepsilon_{\bs{p}}$ being the energy of the chiral fermion.
The Lorentz symmetry requires that the dispersion relation of the chiral fermion has the magnetic moment as \cite{Son:2012zy,Manuel:2013zaa,Chen:2014cla}
\begin{align}
\label{eq:Dispersion_relation_chiral_magnetic_moment}
    \varepsilon_{\bs{p}}
    =|\bs{p}|(1-e\bs{B}\cdot\bs{\Omega}_{\lambda})\,.
\end{align}

Since the invariant phase space is $\sqrt{G}\dif^3\bs{x}\dif^3\bs{p}/(2\pi)^3$, the number density and the number current are written as
\begin{align}
    n_{\lambda}(t,\bs{x})
    &=
    \int\fr{\dif^3\bs{p}}{(2\pi)^3}(1+e\bs{B}\cdot\bs{\Omega}_{\lambda})f_{\lambda}
    +(\text{antifermion})\,,
\end{align}
and
\begin{align}
\label{eq:Current_CKT}
    \bs{j}_{\lambda}(t,\bs{x})
    &=
    \int\fr{\dif^3\bs{p}}{(2\pi)^3}
    \lb[\tilde{\bs{v}}
    +(\tilde{\bs{v}}\cdot\bs{\Omega}_{\lambda})e\bs{B}
    -\varepsilon_{\bs{p}}\bs{\Omega}_{\lambda}\times\fr{\del}{\del\bs{x}}\rb]f_{\lambda}
    +e\bs{E}\times\bs{\sigma}_{\lambda}
    +(\text{antifermion})\,,
\end{align}
where we defined
\begin{align}
    \bs{\sigma}_{\lambda}
    \equiv
    \int\fr{\dif^3\bs{p}}{(2\pi)^3}\bs{\Omega}_{\lambda}f_{\lambda}\,.
\end{align}
Multiplying Eq.~(\ref{eq:Boltzmann_collisionless}) by $\sqrt{G}$ and using Eqs.~(\ref{eq:EoM_x}) and (\ref{eq:EoM_p}), we can reproduce the chiral anomaly
\begin{align}
\label{eq:Chiral_anomaly}
    \del_t n_{\lambda}
    +\bs{\nabla}\cdot\bs{j}_{\lambda}
    &=
    -e^2\int\fr{\dif^3\bs{p}}{(2\pi)^3}\lb(\bs{\Omega}_{\lambda}\cdot\fr{\del}{\del\bs{p}}[f_{\lambda}+\bar{f}_{\lambda}]\rb)\bs{E}\cdot\bs{B}
    \nom
    &=
    \lambda\fr{e^2}{4\pi^2}\bs{E}\cdot\bs{B}\,,
\end{align}
where we used $\bs{\nabla}_{p}\cdot\bs{\Omega}_{\lambda}=2\pi\lambda\delta(\bs{p})$.
At low temperatures, antifermion contributions are suppressed, and we have $f_{\lambda}(\bs{p}=0)=1$ \cite{Son:2012zy}.
Conversely, at high temperatures, antifermions must be included in order to correctly reproduce the chiral anomaly \cite{Manuel:2013zaa}.
This fact is also examined using the Wigner function \cite{Gao:2019zhk}.

\subsection{Transport coefficients}
We derive the current for the relativistic collisionless plasma within the framework of the chiral kinetic theory.
Our derivation partially follows Refs.~\cite{Son:2012zy,Manuel:2013zaa}.
We assume the counting of the gauge field, coordinate derivative, coupling constant, and the chiral chemical potential as $A^\mu=\mcl{O}(\epsilon)$, $\del^{x}_{\mu}=\mcl{O}(\delta)$, $e=\mcl{O}(\delta)$, and $\mu_{5}\equiv\sum_{\lambda=\pm}\lambda\mu_{\lambda}/2=\mcl{O}(\delta^3\epsilon^2)$ respectively.
The assumption that the derivative and the coupling constant can be counted at the same order
is justified a posteriori from the resulting dispersion relation.
The power counting of the chiral chemical potential is taken to be consistent with the chiral anomaly discussed later.

We now solve the Boltzmann equation (\ref{eq:Boltzmann_collisionless}) by expanding the distribution function as%
\footnote{There may exist a contribution $f_{\lambda}^{(\delta^3\epsilon^2)}$ arising from quantum effects.
However, for the O-mode, any current of even order in $\epsilon$ (or equivalently gauge fields) must vanish, since the permittivity depends only on even powers of the background magnetic field, as shown in Sect.~\ref{sec:EM_mag}.}
\begin{align}
    f_{\lambda}(t,\bs{x},\bs{p})
    &=
    \tilde{f}^{(0)}_{\lambda}
    +f^{(\delta\epsilon)}_{\lambda}
    +f^{(\delta^2\epsilon)}_{\lambda}
    +f_{\lambda}^{(\delta^2\epsilon^2)}
    +f_{\lambda}^{(\delta^3\epsilon^3)}\,,
\end{align}
where we defined
\begin{align}
    \tilde{f}^{(0)}_{\lambda}(\varepsilon_{\bs{p}},\bs{p})\equiv\fr{1}{\e^{\beta(\varepsilon_{\bs{p}}-\mu_{\lambda})}+1}\,,
\end{align}
with $\beta$ being the inverse temperature.
Due to the magnetic moment correction in the dispersion relation~(\ref{eq:Dispersion_relation_chiral_magnetic_moment}), the equilibrium distribution function includes a term of order $\mcl{O}(\delta\epsilon)$ as
\begin{align}
    \tilde{f}_{\lambda}^{(0)}(\varepsilon_{\bs{p}},\bs{p})
    &=
    f_{\lambda}^{(0)}(\bs{p})
    -e\bs{B}\cdot\bs{\Omega}_{\lambda}\fr{\del f_{\lambda}^{(0)}(\bs{p})}{\del|\bs{p}|}\,,
\end{align}
where
\begin{align}
    f^{(0)}_{\lambda}(\bs{p})\equiv\fr{1}{\e^{\beta(|\bs{p}|-\mu_{\lambda})}+1}\,.
\end{align}
In the following, we perform the Fourier transformation as $(t,\bs{x})\to(\omega,\bs{k})$.
At order $\mcl{O}(\delta\epsilon)$, the kinetic equation leads to
\begin{align}
    f^{(\delta\epsilon)}_{\lambda}(\omega,\bs{k},\bs{p})
    =
    -\im e\fr{\bs{v}\cdot\bs{E}}{v\cdot k+\im\eta}\fr{\del f^{(0)}_{\lambda}}{\del|\bs{p}|}\,,
\end{align}
where $v^{\mu}\equiv(1,\bs{v})$ and an infinitesimal positive constant $\eta$ is introduced to ensure causality.
The distribution function of order $\mcl{O}(\delta^2\epsilon)$ arises from the Berry curvature and is given by
\begin{align}
    f_{\lambda}^{(\delta^2\epsilon)}(\omega,\bs{k},\bs{p})
    &=
    \lambda e\fr{\omega}{2|\bs{p}|}\fr{\bs{v}\cdot\bs{B}}{v\cdot k+\im\eta}\fr{\del f^{(0)}_{\lambda}}{\del|\bs{p}|}\,.
\end{align}
As we will see, this quantum contribution leads to the permittivity $\varepsilon_{zz} \propto B_{\rm ex}^2$.
Meanwhile, even in the classical theory, the current $\delta\bs{j}\propto B_{\rm ex}^2\delta\bs{E}$ is induced by the distribution function $f_{\lambda}^{(\delta^3\epsilon^3)}$ and results in a correction proportional to the square of the magnetic field.
Thus, we derive the corresponding distribution function by focusing only on the Lorentz force.
At order $\mcl{O}(\delta^2\epsilon^2)$, we obtain
\begin{align}
    f_{\lambda}^{(\delta^2\epsilon^2)}(\omega,\bs{k},\bs{p})
    &=
    -\im e\fr{(\bs{v}\times\bs{B}_{\rm ex})}{v\cdot k+\im\eta}\cdot\fr{\del f_{\lambda}^{(\delta\epsilon)}}{\del \bs{p}}
    \nom
    &=
    -e^2\fr{(\bs{v}\cdot\bs{E})[\bs{v}\cdot(\bs{B}_{\rm ex}\times\bs{k})]}{|\bs{p}|(v\cdot k+\im\eta)^3}\fr{\del f^{(0)}_{\lambda}}{\del|\bs{p}|}\,,
\end{align}
where we used $\bs{B}_{\rm ex}\parallel\bs{E}$ and $(\bs{v}\times\bs{B}_{\rm ex})\cdot\bs{k}=\bs{v}\cdot(\bs{B}_{\rm ex}\times\bs{k})$.
In the same way, we can derive the distribution function of order $\mcl{O}(\delta^3\epsilon^3)$ as
\begin{align}
    f_{\lambda}^{(\delta^3\epsilon^3)}(\omega,\bs{k},\bs{p})
    &=
    \im e^3\fr{\bs{v}\cdot\bs{E}}{|\bs{p}|^2}
    \lb[\fr{(\bs{v}\times\bs{B}_{\rm ex})\cdot(\bs{B}_{\rm ex}\times\bs{k})}{(v\cdot k+\im\eta)^4}+3\fr{[\bs{v}\cdot(\bs{B}_{\rm ex}\times\bs{k})]^2}{(v\cdot k+\im\eta)^5}\rb]
    \fr{\del f^{(0)}_{\lambda}}{\del|\bs{p}|}\,.
\end{align}

Once the distribution functions are obtained, we derive the current by integrating them in momentum space.
Here, we focus on the $z$-component since only the permittivity tensor $\varepsilon_{zz}$ contributes to the O-mode.
We first consider the current $j_{\lambda z}^{(\delta\epsilon)}$.
The integration over $|\bs{p}|$ can be performed using the result from the appendix of Ref.~\cite{Loganayagam:2012pz} as
\begin{align}
    \int\fr{\dif|\bs{p}|}{2\pi^2}|\bs{p}|^2\lb(-\fr{\del}{\del|\bs{p}|}\lb[f_{\lambda}^{(0)}+\bar{f}_{\lambda}^{(0)}\rb]\rb)
    =
    \fr{\mu_{\lambda}^2}{2\pi^2}+\fr{T^2}{6}\,.
\end{align}
The remaining angular integration (see Appendix \ref{app:Angular_int}) leads to
\begin{align}
    j^{(\delta\epsilon)}_{\lambda z}(\omega,\bs{k})
    &=
    \int\fr{\dif^3\bs{p}}{(2\pi)^3}v_zf^{(\delta\epsilon)}_{\lambda}+(\text{antifermion})
    \nom
    &=
    \im\fr{m_{{\rm D}\lambda}^2}{e\omega}F(\xi)E_z\,,
\end{align}
where we defined the parameter $\xi\equiv\omega/|\bs{k}|$ and the dimensionless functions,
\begin{align}
    &F(\xi)
    \equiv
    \fr{1}{2}\lb[\xi^2+(1-\xi^2)L(\xi)\rb]\,,
    \\
\label{eq:Function_L}
    &L(\xi)
    \equiv\int\fr{\dif^2\bs{v}}{4\pi}\fr{\omega}{v\cdot k+\im\eta}
    =
    \fr{\xi}{2}\ln\lb|\fr{1+\xi}{1-\xi}\rb|-\im\fr{\pi\xi}{2}\theta(1-\xi)\,.
\end{align}
The Debye screening mass for the chirality $\lambda$ is given by
\begin{align}
    m_{{\rm D}\lambda}^2
    \equiv
    e^2\lb(\fr{\mu_{\lambda}^2}{2\pi^2}+\fr{T^2}{6}\rb)\,.
\end{align}
Following Ref.~\cite{Son:2012zy,Manuel:2013zaa}, the current of order $\mcl{O}(\delta^2\epsilon)$, namely the CME, is obtained as
\begin{align}
    j_{\lambda z}^{(\delta^2\epsilon)}(\omega,\bs{k})
    &=
    \int\fr{\dif^3\bs{p}}{(2\pi)^3}
    \lb[v_zf^{(\delta^2\epsilon)}_{\lambda}+(\bs{v}\cdot\bs{\Omega}_{\lambda})eB_zf^{(0)}_{\lambda}-\im|\bs{p}|(\bs{\Omega}_{\lambda}\times\bs{k})_{z}f^{(\delta\epsilon)}_{\lambda}\rb]
    +(\text{antifermion})
    \nom
    &=
    \lambda\fr{e\mu_{\lambda}}{4\pi^2}G(\xi)B_{z}\,,
\end{align}
where the dimensionless function $G(\xi)$ is defined as
\begin{align}
    G(\xi)\equiv(1-\xi^2)[1-L(\xi)]\,.
\end{align}
Since the permittivity tensor $\varepsilon_{zz}$ must be even in magnetic fields, the current of order $\mcl{O}(\delta^2\epsilon^2)$ does not appear.
In fact, one can show that $j_{\lambda z}^{(\delta^2\epsilon^2)}=0$ by the explicit calculation. 
We then finally derive the current $j_{\lambda z}^{(\delta^3\epsilon^3)}$.
Writing the components explicitly, we have
\begin{align}
    j_{\lambda z}^{(\delta^3\epsilon^3)}(\omega,\bs{k})
    &=
    -\im e^3\int\fr{\dif^3\bs{p}}{(2\pi)^3}
    \fr{1}{|\bs{p}|^2}\fr{\del f^{(0)}_{\lambda}}{\del|\bs{p}|}
    \lb[\fr{v_xv_z^2}{(v\cdot k+\im\eta)^4}
    -3\fr{v_y^2v_z^2k_x}{(v\cdot k+\im\eta)^5}\rb]k_xB_{\rm ex}^2E_z
    \nom
        &\hspace{3em}+(\text{antifermion})
    \nom
    &=
    \im\fr{e^3B_{\rm ex}^2}{2\pi^2\omega|k_x|^2}I(\xi)E_{z}\,,
\end{align}
where the function $I(\xi)$ is introduced as
\begin{align}
    I(\xi)
    \equiv
    \xi\int\fr{\dif^2\bs{v}}{4\pi}\lb[\fr{v_xv_z^2}{(\xi-v_x+\im\eta)^4}-3\fr{v_y^2v_z^2}{(\xi-v_x+\im\eta)^5}\rb]\,.
\end{align}

\subsection{Dispersion relation in the weak magnetic field limit}
We now examine the dispersion relation of the O-mode in a relativistic plasma.
Combining the linear response relation (\ref{eq:Permittivity_current_general}) with the dispersion relation (\ref{eq:Dispersion_relation_general}), we obtain
\begin{align}
\label{eq:Permittivity_current_z}
    (r_x^2-1)\delta E_z
    =
    \im\fr{1}{\omega}e\delta j_z\,.
\end{align}
In addition to the electromagnetic wave, we assume the fluctuation of the chiral charge density in chirally symmetric matter as
\begin{align}
    n_{5}
    \equiv
    \sum_{\lambda=\pm}\lambda n_{\lambda}
    =\delta n_{5}\,,
\end{align}
while the total number density $n\equiv\sum_{\lambda=\pm}n_{\lambda}$ is assumed not to fluctuate.
Even if the number density fluctuation was included, the result would remain unchanged because it is decoupled within the linear analysis of the fluctuations.
In our counting, the chiral charge density is counted as $n_5\propto \mu_5=\mcl{O}(\delta^3\epsilon^2)$.

Using the results in the previous subsection together with the electric field and the chiral charge density in the form of $\delta E_{z}(t,x)=\delta E_{z}(\omega,k_x)\e^{-\im(\omega t-k_xx)}\,,~\delta n_{5}(t,x)=\delta n_{5}(\omega,k_x)\e^{-\im(\omega t-k_xx)}$, we obtain the fluctuation of the total electric current as
\begin{align}
\label{eq:Current_fluctuation_z}
    e\delta j_{z}(\omega,k_x)
    &=
    e\sum_{\lambda=\pm}
    \lb[\delta j_{\lambda z}
    ^{(\delta\epsilon)}
    +\delta j_{\lambda z}^{(\delta^2\epsilon)}
    +\delta j_{\lambda z}^{(\delta^3\epsilon^3)}\rb]
    \nom
    &=
    \im\fr{m_{\rm D}^2}{\omega}F(\xi)\delta E_z
    +\fr{e^2B_{\rm ex}}{2\pi^2\chi}G(\xi)\delta n_5
    +\im\fr{e^4B_{\rm ex}^2}{\pi^2\omega|k_x|^2}I(\xi)\delta E_{z}\,.
\end{align}
In a chirally symmetric system, the susceptibility of the chiral charge satisfies 
\begin{align}
    \chi\equiv\fr{\del n_5}{\del\mu_5}=\fr{\del n}{\del\mu}\,,
\end{align}
where the chemical potential is defined as $\mu\equiv\sum_{\lambda=\pm}\mu_{\lambda}/2$.

To close the equations, we also need the continuity equation describing the dynamics of the chiral charge.
The anomaly relation for the chiral fermions (\ref{eq:Chiral_anomaly}) yields
\begin{align}
\label{eq:Chiral_anomaly_fluctuation}
    \omega\delta n_5
    =
    \im\fr{e^2B_{\rm ex}}{2\pi^2}\delta E_z\,.
\end{align}
Note that the spatial derivative term in Eq.~(\ref{eq:Chiral_anomaly}) vanishes since there is no current parallel to $k_x$ in our setup.%
\footnote{In general, we also have the chiral electric separation effect (CESE) \cite{Huang:2013iia}, giving a nonlinear contribution as $\delta n_5\delta E_z$. However, there is no such term in the $x$-component of the axial current in our configuration.}
Combining Eqs.~(\ref{eq:Permittivity_current_z}), (\ref{eq:Current_fluctuation_z}), and (\ref{eq:Chiral_anomaly_fluctuation}), we obtain
\renewcommand{\arraystretch}{2}
\begin{align}
    \lb(
    \begin{array}{cc}
        \displaystyle r_x^2-1+\fr{m_{\rm D}^2}{\omega^2}F(\xi)
        +\fr{e^4B_{\rm ex}^2}{\pi^2\omega^2|k_x|^2}I(\xi)& \displaystyle -\im\fr{e^2 B_{\rm ex}}{2\pi^2\chi\omega}G(\xi) \\
        \displaystyle -\im\fr{e^2 B_{\rm ex}}{2\pi^2} & \displaystyle \omega
    \end{array}
    \rb)
    \lb(
    \begin{array}{c}
        \delta E_z \\
        \delta n_5  
    \end{array}
    \rb)
    =0\,.
\end{align}
The eigenvalue of this equation leads to the O-mode dispersion relation including the photon self-energy $\Pi_{\rm O}$.
Therefore, our main result is given by
\begin{align}
\label{eq:Dispersion_relation_anomalous_general}
    \omega^2
    &=
    |k_x|^2
    +\Pi_{\rm O}(\xi)\,,
    \qquad
    \Pi_{\rm O}(\xi)
    \equiv
    m_{\rm D}^2F(\xi)
    +\fr{e^4B_{\rm ex}^2}{\pi^2|k_x|^2}I(\xi)
    +m_{\rm a}^2G(\xi)\,,
\end{align}
where we defined ``anomalous screening mass'' analogous to the Debye screening mass as
\begin{align}
\label{eq:Anomalous_screening_mass}
    m_{\rm a}^2
    \equiv
    \fr{e^4B_{\rm ex}^2}{4\pi^4\chi}\,.
\end{align}
The second and third terms in the photon self-energy are proportional to $B_{\rm ex}^2$ in agreement with the argument of symmetry in Sect.~\ref{sec:EM_mag}.
Eliminating $\delta n_{5}$ in Eq.~(\ref{eq:Current_fluctuation_z}) by using Eq.~(\ref{eq:Chiral_anomaly_fluctuation}), we can obtain the magnetoresistance.
This is an extension of the result in Ref.~\cite{Son:2012bg} for general frequency and wavenumber, including the classical magnetic contribution.

To obtain the analytic expression of the dispersion relation, we consider appropriate limiting cases.
We first take the quasi-static limit $(\xi\ll1)$.
Although the parameter $\xi$ is small, the expansion performed here is not based on the counting of $\delta$ and $\epsilon$ used previously.
Expanding the dimensionless functions up to order $\mcl{O}(\xi)$ (see also Appendix~\ref{app:Expansion_functions}), the photon self-energy becomes
\begin{align}
\label{eq:Photon_self-energy_O-mode_static}
    \Pi_{\rm O}(\xi)
    \simeq
    -\im\fr{\pi m_{\rm D}^2}{4}\xi
    -\im\fr{e^4B_{\rm ex}^2}{16\pi|k_x|^2}\xi
    +m_{\rm a}^2
    +\im\fr{\pi m_{\rm a}^2}{2}\xi\,.
\end{align}
The self-energy retains a real part even in the static limit $(\omega\to0)$.
This is in contrast to the classical result, in which the transverse photon
is gapless.
In the weak magnetic field limit, however, the strictly static limit
$\omega=0$ cannot be taken, because the anomalous mass term is of higher order
than the coordinate derivative.
As a result, the gap appears only at the nonzero frequency,
and we refer to this phenomenon as ``anomalous dynamical screening.''
Note that the X-mode is not
affected by the anomaly and remains gapless; consequently, dynamical magnetic
fields are \emph{partially} screened.

Next, we examine the quasi-long-wavelength limit $(\xi\gg1)$.
Using the expansion of the dimensionless functions in Appendix \ref{app:Expansion_functions}, we have
\begin{align}
\label{eq:Photon_self-energy_O-mode_long}
    \Pi_{\rm O}(\xi)
    \simeq
    \omega_{\rm p}^2
    +\omega_{\rm a}^2
    +\fr{\omega_{\rm p}^2}{5}\fr{1}{\xi^2}
    -\fr{2\omega_{\rm a}^2}{5}\fr{1}{\xi^2}\,,
\end{align}
where $\omega_{\rm p}\equiv m_{\rm D}/\sqrt{3}$ is the plasma frequency and $\omega_{\rm a}\equiv m_{\rm a}/\sqrt{3}$ is the ``anomalous plasma frequency.'' 
In contrast to the quasi-static limit, the classical magnetic correction is not included up to order $\mcl{O}(1/\xi^2)$.
This result is consistent with the the well-known behavior that the O-mode is not affected by the background magnetic field in the long-wavelength limit $(\bs{k}\to0)$.
Thus, the magnetic dependence comes only from the anomaly.
The dispersion relation of the plasma oscillation is then written as
\begin{align}
\label{eq:Dispersion_relation_anomalous_plasma_osillation}
    \omega^2
    &\simeq
    \omega_{\rm p}^2(1+\zeta^2)
    +\fr{6}{5}|k_x|^2\lb(1-\fr{4}{3}\zeta^2\rb)\,,
\end{align}
where $\zeta\equiv m_{\rm a}/m_{\rm D}=\omega_{\rm a}/\omega_{\rm D}$
.
From this expression, we can find that the chiral anomaly gives rise to the correction to both the gap and the velocity of the plasma oscillation.

Let us comment on the effect of fermion masses.
While we have used the chiral kinetic theory, real fermions such as electrons and quarks are massive, and this limits the applicability of our discussion.
Although the CME is unaffected by fermion masses \cite{Fukushima:2008xe}, the anomaly equation (\ref{eq:Chiral_anomaly_fluctuation}) is modified through the explicit breaking of the U(1)$_{\rm A}$ symmetry.%
\footnote{Unlike the CME, the CSE is affected by fermion masses \cite{Metlitski:2005pr,Gorbar:2013upa,Guo:2016dnm}. However, since the anomaly relation (\ref{eq:Chiral_anomaly_fluctuation}) does not include the axial current in our setup, our result is independent of the CSE.}
This mass effect induces the chirality flipping, which violates chiral charge conservation.
In the collisionless regime, however, scattering is negligible, so our results remain valid.
In Sect.~\ref{sec:Imp_neutron_star}, we estimate the chirality flipping time in neutron star matter, which turns out to be sufficiently longer than the relaxation time.

\subsection{Dispersion relation in the strong magnetic field limit}
So far, we have focused on the weak magnetic field regime using the chiral kinetic theory.
In this subsection, we turn to the strong magnetic field regime, in which the fermions occupy only the LLL.
In contrast to the chiral kinetic theory, the background magnetic field is taken to be of order unity $B_{\rm ex}=\mcl{O}(1)$, while the dynamical electromagnetic fields are counted as $\delta\bs{E},\delta\bs{B}=\mcl{O}(\delta\epsilon)$.
Accordingly, we adopt the power counting as $\delta A^{\mu}=\mcl{O}(\epsilon), \del^x_\mu=\mcl{O}(\delta)$, $e=\mcl{O}(\delta^{2/3})$, and $\mu_5=\mcl{O}(\delta^{4/3}\epsilon)$, where $\delta A^{\mu}$ is the gauge field fluctuation.
The coupling constant and the chiral chemical potential are counted so as to be compatible with the derived dispersion relation and the chiral anomaly, respectively.

Since the dispersion relation of the LLL is $\varepsilon_{\lambda}=\lambda p_z$, we can treat the motion of the chiral fermions in (1+1)-dimensional spacetime effectively.
The Boltzmann equation takes the form
\begin{align}
\label{eq:Boltzmann_eq_LLL}
    \fr{\del f_{\lambda}}{\del t}
    +\dot{z}\fr{\del f_{\lambda}}{\del z}
    +\dot{p}_z\fr{\del f_{\lambda}}{\del p_z}
    =
    0\,,
\end{align}
where the velocity and the equation of motion for the fermions are given by
\begin{align}
    \dot{z}
    =
    \lambda\,,
    \qquad
    \dot{p}_z
    =
    eE_z\,.
\end{align}
The transverse degrees of freedom are replaced by the Landau degeneracy $|eB_{\rm ex}|/(2\pi)$.

Similarly to the case of the chiral kinetic theory, we can solve the Boltzmann equation by expanding the distribution function as
\begin{align}
    f_{\lambda}(t,z,p_z)
    &=
    f^{(0)}_{\lambda}
    +f^{(\delta\epsilon)}_{\lambda}\,,
\end{align}
where we defined
\begin{align}
    f^{(0)}_{\lambda}(p_z)\equiv\fr{1}{\e^{\beta(|p_z|-\mu_{\lambda})}+1}\,.
\end{align}
The distribution function of order $\mcl{O}(\delta\epsilon)$ is derived as
\begin{align}
    f_{\lambda}^{(\delta\epsilon)}(\omega,k_z,p_z)
    &=
    -\im\fr{eE_z}{\omega-\lambda k_z+\im\eta}\fr{\del f_{\lambda}^{(0)}}{\del p_z}\,.
\end{align}

The current of order $\mcl{O}(\delta)$ can be obtained as
\begin{align}
    j_{\lambda z}^{(\delta)}
    &=
    \fr{|eB_{\rm ex}|}{2\pi}\int\fr{\dif p_z}{2\pi}\dot{z}f_{\lambda}^{(0)}+(\text{antifermion})
    \nom
    &=
    \lambda\fr{\mu_{\lambda}}{4\pi^2}|eB_{\rm ex}|\,.
\end{align}
At order $\mcl{O}(\delta^2\epsilon)$, the current can be written as
\begin{align}
    j_{\lambda z}^{(\delta^2\epsilon)}(\omega,k_z)
    &=
    \fr{|eB_{\rm ex}|}{2\pi}\int\fr{\dif p_z}{2\pi}\dot{z}f_{\lambda}^{(\delta\epsilon)}+(\text{antifermion})
    \nom
    &=
    \im\fr{|eB_{\rm ex}|}{2\pi^2}\fr{k_z}{\omega^2-k_z^2}eE_z\,.
\end{align}
In our setup, we set $k_z=0$, and  only the current $j_{\lambda z}^{(\delta)}$ remains.
Thus, the fluctuation of the total electric current becomes
\begin{align}
\label{eq:Current_fluctuation_LLL}
    e\delta j_z
    &=
    e\sum_{\lambda=\pm}\delta j_{\lambda z}^{(\delta)}
    =
    \fr{e|eB_{\rm ex}|}{2\pi^2\chi}\delta n_5\,.
\end{align}
The counting of the chiral charge density is the same as that of the chiral chemical potential $\delta n_5=\mcl{O}(\delta^{4/3}\epsilon)$.

From Eqs.~(\ref{eq:Permittivity_current_z}), (\ref{eq:Chiral_anomaly_fluctuation}), and (\ref{eq:Current_fluctuation_LLL}), we obtain the dispersion relation
\begin{align}
\label{eq:Dispersion_relation_LLL}
    \omega^2
    &=
    |k_x|^2+\Pi_{\rm O}\,,
    \qquad
    \Pi_{\rm O}=m_{\rm a}^2\,.
\end{align}
Since the susceptibility is $\chi\sim|eB_{\rm ex}|$, the anomalous screening mass becomes $m_{\rm a}\sim e^{3/2}\sqrt{B_{\rm ex}}$, giving rise to the soft scale.%
\footnote{In the LLL approximation, the system effectively becomes (1+1)-dimensional. Thus, the symmetry argument presented in Sect.~\ref{sec:EM_mag} is no longer valid.}
This is consistent with the derivative expansion.
We thus find that the anomalous screening emerges even in the strong magnetic field limit.
In contrast to the weak field case, however, the static limit can be taken in this regime, and the O-mode is genuinely screened.

\section{Implications for neutron star phenomenology}
\label{sec:Imp_neutron_star}
So far, we have considered a general relativistic plasma.
In this section, as an example, we focus on degenerate electron matter in the core of neutron stars and discuss some implications.
We assume that the electron chemical potential is much larger than both the temperature and the electron mass $(\mu_{\rm e}\gg T, m_{\rm e})$.

\subsection{Relaxation time}
The screening characterizes the infrared (IR) scale of scattering processes.
We then discuss the effect of the anomalous dynamical screening on the relaxation time.
Here, we focus on the electron--electron scattering as a photon-mediated process.
Indeed, the neutron superfluid emerges at temperature $T\lesssim10^{9}~{\rm K}$~\cite{Page:2010aw}, where the electron--electron scattering becomes the dominant process.
The scattering amplitude $M$ includes the photon propagator as
\begin{align}
    M
    =
    \fr{C_1}{\omega^2-|\bs{k}|^2-\Pi_{\rm L}}
    +\fr{C_2}{\omega^2-|\bs{k}|^2-\Pi_{\rm T}}\,,
\end{align}
where $\Pi_{\rm L},\Pi_{\rm T}$ are the longitudinal and transverse components of the photon self-energy.
The details of the coefficients $C_1, C_2$ are unimportant in this discussion.
The scattering is dominated by the IR cutoff of momentum $(k_{\rm IR}^2\sim \Pi_{\rm L}, \Pi_{\rm T})$.
In general, due to the presence of the background magnetic field, these are further decomposed into more components.
However, we focus only on the O-mode and set $\Pi_{\rm T}=\Pi_{\rm O}, |\bs{k}|=|k_x|$ in the following discussion.
Due to the Pauli blocking, only fermions near the Fermi surface contribute to the scattering, and the typical energy transfer is of the order of the temperature, $\omega\sim T$.
Therefore, it is sufficient to use the photon self-energy in the quasi-static limit.
A simple dimensional analysis then leads to the following parameter dependence of the relaxation time: 
\begin{align}
    \fr{1}{\tau}\sim\alpha^2\fr{T^2}{k_{\rm IR}}\,,
\end{align}
where $\alpha=e^2/(4\pi)$ is the fine structure constant.
However, the Landau damping can modify the temperature behavior of the relaxation time, meaning that the non-Fermi liquid arises \cite{Heiselberg:1993cr}.
Instead of tracing the detailed discussion in Ref.~\cite{Heiselberg:1993cr}, we provide a schematic argument by focusing on the parameters.
When the background magnetic field is absent, the photon self-energy (\ref{eq:Photon_self-energy_O-mode_static}) becomes $\Pi_{\rm O}\simeq-\im\fr{\pi m_{\rm D}^2}{4}\xi$, resulting in the classical IR cutoff as $k_{\rm IR}\sim(m_{\rm D}^2\omega)^{1/3}\equiv k_{\rm c}$.
From this expression, we can reproduce the conventional parameter dependence of the relaxation time presented in Ref.~\cite{Heiselberg:1993cr} as 
\begin{align}
\label{eq:Relaxation_time_Landau}
    \fr{1}{\tau_{\rm c}}
    \sim
    \alpha^2\fr{T^2}{k_{\rm c}}
    \sim
    \alpha^2\fr{T^{5/3}}{m_{\rm D}^{2/3}}\,.
\end{align}

We now turn on a weak background magnetic field and examine the resulting correction.
Writing the IR cutoff as $k_{\rm IR}\sim k_{\rm c}+\delta k$, we solve $k_{\rm IR}^2\sim\Pi_{\rm O}(k_{\rm IR})$ using Eq.~(\ref{eq:Photon_self-energy_O-mode_static}) up 
to linear order in $\delta k$.
The resulting momentum is
\begin{align}
    k_{\rm IR}
    \sim
    k_{\rm c}\lb[1+\zeta^2+\lb(\fr{e^2B_{\rm ex}^2}{m_{\rm D}^4}+\zeta^2\rb)\fr{m_{\rm D}^2}{k_{\rm c}^2}\rb]\,.
\end{align}
Here, the numerical coefficients, which are not essential for the present discussion, are omitted.
The relaxation time in the weak magnetic field limit can be parametrically expressed as
\begin{align}
\label{eq:Relaxation_time_weak_field}
    \fr{1}{\tau_{\rm w}}
    \sim
    \alpha^2\fr{T^{5/3}}{m_{\rm D}^{2/3}}(1+\zeta^2)
    +\alpha^2 T\lb(\fr{e^2B_{\rm ex}^2}{m_{\rm D}^4}+\zeta^2\rb)\,.
\end{align}
The quantum correction is small $(\zeta\ll1)$ and does not substantially affect the relaxation time.
Although the magnetic field modifies the temperature dependence, the leading behavior remains proportional to $T^{5/3}$.

On the other hand, the result in the strong magnetic field (\ref{eq:Dispersion_relation_LLL}) leads to the IR cutoff $k_{\rm IR}\sim m_{\rm a}$.
Therefore, the relaxation time is characterized by the anomalous screening mass as
\begin{align}
\label{eq:Relaxation_time_strong_field}
    \fr{1}{\tau_{\rm s}}
    \sim
    \alpha^2\fr{T^2}{m_{\rm a}}\,.
\end{align}
From the above, we can expect that the temperature dependence of the relaxation time changes as the background magnetic field increases.

The relaxation time governs the dissipative transport.
Among them, we focus on the shear viscosity.
The shear viscosity of neutron stars was first derived in Refs.~\cite{Flowers:1979,Cutler:1987}, and the result is modified by the dynamical screening, which is more dominant~\cite{Shternin:2008ia,Shternin:2008es}.
The shear viscosity is given by $\eta=\fr{1}{5}n_{\rm e}\mu_{\rm e}\tau$.
To investigate the effect of the anomalous screening, we evaluate the ratio between Eqs.~(\ref{eq:Relaxation_time_Landau}) and (\ref{eq:Relaxation_time_strong_field}):
\begin{align}
\label{eq:Relaxation_time_ratio}
    \fr{\tau_{\rm s}}{\tau_{\rm c}}
    \sim
    \fr{m_{\rm a}}{m_{\rm D}^{2/3}T^{1/3}}
    \sim
    10~\lb(\fr{T}{10^{6}~{\rm K}}\rb)^{-1/3}
    \lb(\fr{\vphantom{T}\mu_{\rm e}}{10^{2}~{\rm MeV}}\rb)^{-2/3}
    \lb(\fr{B_{\rm ex}}{10^{18}~{\rm G}}\rb)^{1/2}\,,
\end{align}
where we substituted the susceptibility of the electron gas in the LLL approximation $\chi_{\rm e}=|eB_{\rm ex}|/(2\pi^2)$.
The magnetic fields of order $10^{18}~{\rm G}$ corresponds to the theoretical upper limit in neutron stars~\cite{Shapiro:1991,Cardall:2000bs,Ferrer:2010wz}.
This indicates that sufficiently strong magnetic fields tend to enhance the shear viscosity, whereas extrapolating our result to the intermediate magnetic field region ($B_{\rm ex} \lesssim 10^{16}~\rm{G}$) suggests that the shear viscosity is instead suppressed.

The shear viscosity plays an important role, e.g., in stellar oscillations.
In a rotating star, a collective excitation can arise due to the Coriolis force.
This is called the r-mode.
Due to the Chandrasekhar--Friedman--Schutz (CFS) mechanism~\cite{Chandrasekhar:1970pjp,Friedman:1978hf}, the r-mode is known to be unstable, emitting gravitational waves~\cite{Andersson:1997xt}.
The r-mode instability is suppressed by the viscosity, and the critical angular velocity is determined by the competition between the timescales of gravitational radiation and viscous damping.
Therefore, in the presence of intermediate magnetic fields, the r-mode instability may become more likely.

\subsection{Screening length}
The presence of the anomalous screening mass implies that a finite energy scale is required to excite dynamical magnetic fluctuations perpendicular to the background magnetic field.
We here focus on the result in the weak field limit~(\ref{eq:Photon_self-energy_O-mode_static}), because the magnetic field strengths accessible in this regime are more relevant for realistic descriptions of neutron stars.
In the framework of the chiral kinetic theory, the anomalous screening mass can appear at finite frequency $(\omega\neq 0)$.
However, we here extrapolate our result to the static limit $(\omega=0)$.
In this case, the $z$-component of the gauge field, which contributes to the O-mode, behaves as
\begin{align}
    A_{z}(x)\sim\e^{-|x|/l_{\rm a}}\,,
\end{align}
where the screening length is defined as $l_{\rm a}\equiv 1/m_{\rm a}$.

We now evaluate the screening length.
In the ideal gas approximation, the electron number susceptibility is given by $\chi_{\rm e}=\mu_{\rm e}^2/\pi^2$.
Substituting this into Eq.~(\ref{eq:Anomalous_screening_mass}), we obtain
\begin{align}
    l_{\rm a}
    =
    \fr{\mu_{\rm e}}{2\alpha B_{\rm ex}}
    \sim
    10^{-9}~{\rm m}~\lb(\fr{B_{\rm ex}}{10^{14}~{\rm G}}\rb)^{-1}\lb(\fr{\vphantom{T}\mu_{\rm e}}{10^{2}~{\rm MeV}}\rb)\,.
\end{align}
The typical size of a neutron star is much larger than the screening length.
Consequently, the magnetic degree of freedom associated with the O-mode becomes irrelevant for macroscopic electromagnetic dynamics.

Magnetohydrodynamics (MHD) assumes that magnetic fields are not screened.
The anomalous dynamical screening may therefore require a modification of the standard MHD description of magnetized relativistic plasmas.
Moreover, the above discussion is not limited to neutron stars and can be applied to other contexts, such as the early Universe.

\subsection{Chirality flipping time}
We finally comment on the chirality flipping induced by the electron mass.
Since the chirality flipping is caused by the longitudinal photon \cite{Schafer:2001za,Hanai:2022yfh}, the dynamical screening does not need to be taken into account.
In this case, the electron-proton scattering is the dominant process, and the chirality flipping time is \cite{Grabowska:2014efa, Dvornikov:2015iua}
\begin{align}
    &\tau_{\rm m}
    \simeq
    \fr{\pi\mu_{\rm e}^3}{\alpha^2 m_{\rm e}^2m_{\rm p}T}\lb(\ln\fr{4\mu_{\rm e}^2}{m_{\rm D}^2}-1\rb)^{-1}
    \sim
    10^{18}~({\rm MeV})^{-1}\lb(\fr{T}{10^6~{\rm K}}\rb)^{-1}\,,
\end{align}
where $m_{\rm e}$ denotes the electron mass.
However, in the parameter region of interest, the chirality flipping time is larger than the relaxation time $(\tau_{\rm m}\gg \tau_{\rm c},\tau_{\rm w},\tau_{\rm s})$.
Thus, in the collisionless regime, the chirality flipping can be ignored.

\section{Conclusion and outlook}
\label{sec:Conclusion_Outlook}
In this paper, we have studied the collective modes of relativistic magnetized plasmas in the collisionless regime, taking the chiral anomaly into account.
We showed that the chiral anomaly induces a modification to the O-mode dispersion relation within both the chiral kinetic theory (\ref{eq:Dispersion_relation_anomalous_general}) and the LLL approximation~(\ref{eq:Dispersion_relation_LLL}).
In the former case, we derived the photon self-energy in two limiting expressions: in the quasi-static limit (\ref{eq:Photon_self-energy_O-mode_static}) and the quasi-long-wavelength limit (\ref{eq:Photon_self-energy_O-mode_long}).
We found that the self-energy has the anomalous screening mass remaining in the static limit, which is distinct from the classical gapless transverse mode.
A similar gap also appears in the LLL approximation.

Due to the anomalous screening, the temperature dependence of the relaxation time in sufficiently strong magnetic fields is different from the conventional behavior by the Landau damping, as shown in Eq.~(\ref{eq:Relaxation_time_strong_field}).
If this is valid even in the intermediate magnetic field region, the resulting viscosity is suppressed, and the r-mode instability can be enhanced.
The presence of the anomalous screening also implies that the conventional MHD description is modified in magnetized relativistic plasmas.

There are several possible directions for future work.
In this paper, we investigated the collective mode in the weak and strong magnetic field limits.
On the other hand, it would be interesting to study collective modes in the intermediate magnetic field regime, which cannot be treated rigorously within our framework.
The anomalous effects may become comparable to or even larger than the classical contributions in this region.
In the context of QCD and Dirac semimetals, it is shown that a quadratic collective mode is induced by the chiral anomaly~\cite{Sogabe:2019gif}.
There, in a phase with broken chiral symmetry, the resulting Nambu--Goldstone mode (e.g., the pion) plays the role of the chiral chemical potential in our setup.
Therefore, we can expect that a similar collective mode arises.

Another important direction is the application to quark matter, such as color superconductivity in the cores of neutron stars and quark-gluon plasma (QGP) in the heavy-ion collision experiments.
In the case of the two-flavor color superconductivity (2SC) \cite{Barrois:1977xd,Bailin:1983bm,Alford:1997zt,Rapp:1997zu}, one of the three color degrees of freedom does not participate in Cooper pairing.
The unpaired quark remains gapless and behaves as a relativistic fermion.
A strong magnetic field of $\sim 10^{18}~{\rm G}$ can also be created in the heavy-ion collision experiments \cite{Kharzeev:2007jp,Skokov:2009qp}.
The chiral anomaly can modify the transport properties in these systems.

\section*{Acknowledgment}
I am grateful to Naoki Yamamoto for helpful discussions and for carefully reading the manuscript.
I also thank Yoshimasa Hidaka for useful comments.

\vspace{0.2cm}
\noindent
\bibliographystyle{ptephy}
\bibliography{Ref}

\appendix

\section{Angular integrals}
\label{app:Angular_int}
In this appendix, we derive the angular integral used in Sect.~\ref{sec:Collective_modes} and summarize the formulae.
We first consider the integral
\begin{align}
    J^{i}(\omega,\bs{k})
    \equiv
    \int\fr{\dif^2\bs{v}}{4\pi}\fr{v^i}{v\cdot k+\im\eta}\,.
\end{align}
Since the only available vector is $k^i$, this integral can be expressed as
\begin{align}
    J^i(\omega,\bs{k})
    =
    k^iA(\omega,\bs{k})\,,
\end{align}
where $A(\omega,\bs{k})$ is a function to be determined.
Multiplying both sides by $k^i$, we have
\begin{align}
    A(\omega,\bs{k})
    &=
    \fr{1}{|\bs{k}|^2}\int\fr{\dif^2\bs{v}}{4\pi}\fr{\bs{v}\cdot\bs{k}}{v\cdot k+\im\eta}
    \nom
    &=
    \fr{1}{|\bs{k}|^2}\int\fr{\dif^2\bs{v}}{4\pi}\lb(-1+\fr{\omega}{v\cdot k+\im\eta}\rb)
    \nom
    &=
    \fr{1}{|\bs{k}|^2}\lb[-1+L(\xi)\rb]\,,
\end{align}
where the function $L(\xi)$ is defined in Eq.~(\ref{eq:Function_L}).

We next compute
\begin{align}
    J^{ij}(\omega,\bs{k})
    \equiv
    \int\fr{\dif^2\bs{v}}{4\pi}\fr{v^iv^j}{v\cdot k+\im\eta}\,.
\end{align}
The available tensors are $\delta^{ij}$ and $k^{i}k^{j}$.
Thus, we can write
\begin{align}
    J^{ij}
    =
    \delta^{ij}B(\omega,\bs{k})+k^ik^jC(\omega,\bs{k})\,.
\end{align}
The products with $\delta^{ij}$ and $k^ik^j$ lead to the following coefficients respectively:
\begin{align}
    &B(\omega,\bs{k})
    =
    \fr{\omega}{2|\bs{k}|^2}\lb[1-\fr{k^2}{\omega^2}L(\xi)\rb]\,,
    \\
    &C(\omega,\bs{k})
    =
    -\fr{\omega}{2|\bs{k}|^4}\lb[1-\fr{k^2}{\omega^2}L(\xi)\rb]
    -\fr{\omega}{|\bs{k}|^4}\lb[1-L(\xi)\rb]\,.
\end{align}
Collecting the results, the angular integrals are summarized as
\begin{align}
    &\int\fr{\dif^2\bs{v}}{4\pi}=1\,,
    \\
    &\int\fr{\dif^2\bs{v}}{4\pi}v^i=0\,,
    \\
    &\int\fr{\dif^2\bs{v}}{4\pi}v^iv^j
    =\fr{1}{3}\delta^{ij}\,,
    \\
    &\int\fr{\dif^2\bs{v}}{4\pi}\fr{\omega}{v\cdot k+\im\eta}
    =
    L\,,
    \\
    &\int\fr{\dif^2\bs{v}}{4\pi}\fr{v^i}{v\cdot k+\im\eta}
    =
    -\fr{k^i}{|\bs{k}|^2}(1-L)\,,
    \\
    &\int\fr{\dif^2\bs{v}}{4\pi}\fr{v^iv^j}{v\cdot k+\im\eta}
    =
    \lb(\delta^{ij}-\fr{k^ik^j}{|\bs{k}|^2}\rb)\fr{\omega}{2|\bs{k}|^2}\lb(1-\fr{k^2}{\omega^2}L\rb)
    -\fr{k^ik^j}{|\bs{k}|^2}\fr{\omega}{|\bs{k}|^2}(1-L)\,.
\end{align}

\section{Expansion of the dimensionless functions}
\label{app:Expansion_functions}
In this appendix, we derive the expressions of the dimensionless functions, $F(\xi)$, $G(\xi)$, and $I(\xi)$ in the quasi-static and quasi-long-wavelength limits.
The function $L(\xi)$ can be expanded as
\begin{align}
    L(\xi)
    \simeq
    \lb\{
    \begin{array}{lc}
        \displaystyle -\im\fr{\pi}{2}\xi+\xi^2+\mcl{O}(\xi^3) & (\xi\ll1) \\
        \displaystyle 1+\fr{1}{3\xi^2}+\fr{1}{5\xi^4}+\mcl{O}(\xi^{-6})& (\xi\gg1)
    \end{array}\rb.\,.
\end{align}
From these expressions, one can easily expand the functions, $F(\xi)$ and $G(\xi)$, as
\begin{align}
    &F(\xi)
    \simeq
    \lb\{
    \begin{array}{lc}
        \displaystyle -\im\fr{\pi}{4}\xi+\mcl{O}(\xi^2) & (\xi\ll1) \\
        \displaystyle \fr{1}{3}+\fr{1}{15\xi^2}+\mcl{O}(\xi^{-4})& (\xi\gg1)
    \end{array}\rb.\,,
    \\
    &G(\xi)
    \simeq
    \lb\{
    \begin{array}{lc}
        \displaystyle 1+\im\fr{\pi}{2}\xi+\mcl{O}(\xi^2) & (\xi\ll1) \\
        \displaystyle \fr{1}{3}-\fr{2}{15\xi^2}+\mcl{O}(\xi^{-4})& (\xi\gg1)
    \end{array}\rb.\,.
\end{align}

We now consider the function $I(\xi)$.
In our setup, it is convenient to express the velocity as $\bs{v}=(\cos\theta, \sin\theta\cos\phi,\sin\theta\sin\phi)$, and we can easily perform the integration over $\phi$ as
\begin{align}
    I(\xi)
    &=
    \fr{1}{4\pi}\xi
    \int_{-1}^{1}\dif v_x\int_{0}^{2\pi}\dif\phi\lb[\fr{v_x\sin^2\theta\sin^2\phi}{(\xi-v_x+\im\eta)^4}
    -3\fr{\sin^4\theta\cos^2\phi\sin^2\phi}{(\xi-v_x+\im\eta)^5}\rb]
    \nom
    &=
    \fr{1}{4}\xi
    \int_{-1}^{1}\dif v_x\lb[\fr{v_x(1-v_x^2)}{(\xi-v_x+\im\eta)^4}-\fr{3}{4}\fr{(1-v_x^2)^2}{(\xi-v_x+\im\eta)^5}\rb]\,.
\end{align}
Rather than evaluating the remaining integral exactly, it is sufficient for our purposes to examine the behavior of $I(\xi)$ in two limiting cases.
In the quasi-static limit, the pole gives rise to an imaginary part, since for $|x|<1$, we have
\begin{align}
    \fr{1}{(x+\im\eta)^{n}}
    =\mcl{P}\fr{1}{x^n}-\im\pi\fr{(-1)^{n-1}}{(n-1)!}\delta^{(n-1)}(x)\,,
\end{align}
where $n$ is a positive integer, $\delta^{(n)}(x)$ denotes the $n$-th derivative of the delta function, and $\mcl{P}$ stands for the principal value.
Up to order $\mcl{O}(\xi)$, the real part vanishes as
\begin{align}
    {\rm Re}[I(\xi)]
    &\simeq
    \fr{1}{4}\xi
    \mcl{P}\int_{-1}^{1}\dif v_x\lb[\fr{v_x(1-v_x^2)}{v_x^4}-\fr{3}{4}\fr{(1-v_x^2)^2}{v_x^5}\rb]
    +\mcl{O}(\xi^2)
    \nom
    &=
    \mcl{O}(\xi^2)\,.
\end{align}
The imaginary part becomes
\begin{align}
    {\rm Im}[I(\xi)]
    &\simeq
    -\fr{\pi}{16}\xi\,.
\end{align}
In the quasi-long-wavelength limit, the integrand does not have any poles, and the function gives just higher order terms as $I(\xi)\simeq\mcl{O}(\xi^{-3})$.
From the above, up to the order of interest, the function $I(\xi)$ can be expanded as
\begin{align}
    I(\xi)\simeq
    \lb\{
    \begin{array}{lc}
        \displaystyle -\im\fr{\pi}{16}\xi+\mcl{O}(\xi^2) & (\xi\ll1) \\
        \displaystyle \mcl{O}(\xi^{-3})& (\xi\gg1)
    \end{array}\rb.\,.
\end{align}

\end{document}